\documentclass[runningheads]{llncs}

\usepackage{amsmath}
\usepackage{bm}
\usepackage{caption}
\usepackage[numbers]{natbib}
\usepackage{url}

\usepackage{subcaption}
\usepackage{amsmath} 
\usepackage[T1]{fontenc}
\usepackage{graphicx}
\usepackage{float}
\begin{document}
\title{Predicting ChatGPT Use in Assignments: Implications for AI-Aware Assessment Design }

%
%
\author{Surajit Das\inst{1}\orcidID{0009-0008-6692-6697}\thanks{The full dataset, code and the complete set of survey questions used in this study may be provided upon reasonable request to the corresponding author} \and
Aleksei Eliseev\inst{2}\orcidID{0000-0002-4549-5356} }
\authorrunning{Surajit Das and 	Aleksei Eliseev}
%

\institute{ITMO University, St. Petersburg, Russia \\ \email{mr.surajitdas@gmail.com}
\and Moscow Institute of Physics and Technology, Moscow, Russia \\ \email{	eliseev.av@mipt.ru}}
\maketitle              
\begin{abstract}
The rise of generative AI tools like ChatGPT has significantly reshaped education, sparking debates about their impact on learning outcomes and academic integrity. While prior research highlights opportunities and risks, there remains a lack of quantitative analysis of student behavior when completing assignments. Understanding how these tools influence real-world academic practices, particularly assignment preparation, is a pressing and timely research priority.

This study addresses this gap by analyzing survey responses from 388 university students, primarily from Russia, including a subset of international participants. Using the XGBoost algorithm, we modeled predictors of ChatGPT usage in academic assignments. Key predictive factors included learning habits, subject preferences, and student attitudes toward AI. Our binary classifier demonstrated strong predictive performance, achieving 80.1\% test accuracy, with 80.2\% sensitivity and 79.9\% specificity. The multiclass classifier achieved 64.5\% test accuracy, 64.6\% weighted precision, and 64.5\% recall, with similar training scores, indicating potential data scarcity challenges.

The study reveals that frequent use of ChatGPT for learning new concepts correlates with potential overreliance, raising concerns about long-term academic independence. These findings suggest that while generative AI can enhance access to knowledge, unchecked reliance may erode critical thinking and originality. We propose discipline-specific guidelines and reimagined assessment strategies to balance innovation with academic rigor. These insights can guide educators and policymakers in ethically and effectively integrating AI into education.

\end{abstract}


\section{Introduction}

The rapid emergence of generative AI tools such as ChatGPT has transformed workflows, decision-making processes, and human-machine interaction across diverse domains. Remarkably, within just two months of its release, ChatGPT amassed over 100 million users worldwide, representing the fastest adoption of any consumer application to date \cite{hu2023}. This unprecedented surge has also significantly impacted education, offering new possibilities for accessible knowledge generation, academic support, and personalized learning. However, it has simultaneously raised concerns about academic integrity, student engagement, and assessment practices \cite{zastudil2023}.

Higher education institutions now face the dual challenge of harnessing the benefits of generative AI while preserving academic standards. Students increasingly use such tools for a variety of academic tasks, including essay writing, programming, and exam preparation. Yet, the motivations behind these behaviors and their implications for learning outcomes remain underexplored, particularly in quantitative terms. Students may leverage generative AI either to deepen understanding or as a shortcut to bypass intellectual effort—two contrasting behaviors with profoundly different implications for education policy and pedagogy. This study quantitatively investigates these motivations, contexts, and behavioral patterns, addressing questions such as: Who uses ChatGPT for academic work? What types of students are more likely to rely on it? And what patterns might signal ethical versus overreliant use?

Although prior research has begun to explore the educational implications of generative AI—such as the impact on learning strategies \cite{bettayeb2024impact}, institutional policy \cite{rejeb2024webmining}, and meta-analyses of learning outcomes \cite{wang2025effect}—few studies provide predictive models grounded in behavioral data from real academic contexts.

To address this gap, we analyze survey responses from 388 university students (predominantly Russian, including international students) and apply Extreme Gradient Boosting (XGBoost) to identify key predictors of ChatGPT usage. Our primary research question is:

\begin{itemize}
    \item What behavioral, demographic, or attitudinal factors predict student use of ChatGPT in academic assignments?
\end{itemize}

By employing an interpretable machine learning model, our goal is to uncover meaningful behavioral predictors of generative AI adoption. These insights can assist educators and policymakers in designing discipline-specific guidelines and assessment strategies that balance innovation with academic integrity.

\section{Related Work}
Generative AI’s emergence has sparked widespread educational inquiry, ranging from implementation strategies to ethical and pedagogical implications. Early reviews highlight the promise and pitfalls of large language models (LLMs) in educational settings. Tan et al. \cite{tan2024} emphasize the urgency of rethinking academic integrity frameworks in the era of ChatGPT, while Shailendra et al. \cite{shailendra2024} propose a structured framework for GenAI adoption that supports both personalization and accountability.

Empirical insights are also growing. Zastudil et al. \cite{zastudil2023} conducted interviews with students and faculty to explore how generative AI affects productivity and critical thinking. Gutiérrez and Strachan \cite{gutierrez2024} describe practical GenAI deployment strategies, while Fleury et al. \cite{fleury2024} summarize real-world classroom adoption experiences. These works advocate for AI-literate, ethically aware curricula.

T4E-specific studies further demonstrate evolving instructional practices. Khwaja et al. \cite{khwaja2023} present a case study on orchestrating active learning in hybrid classrooms using EdTech, offering practical insights into student engagement. Other T4E proceedings emphasize the importance of aligning AI tools with course outcomes and learner behaviors \cite{edu2024}.

Despite these advances, quantitative modeling of why and when students choose to use AI—especially in real academic tasks such as assignment preparation—remains limited. Our study fills this gap by analyzing behavioral, disciplinary, and attitudinal predictors of ChatGPT usage using interpretable machine learning, offering data-driven recommendations for educators and EdTech policy designers.

\section{General Discussion}

 For the study purpose, a self reported survey was carried out amid university students to capture different types of variables pertaining to demography, daily practice, engagements, etc. which possibly shape AI engagement in academics. Several questions were setup to capture the responses which were validated by the experts and the inconstant responses were eliminated. However, we presume that the dataset is not absolute noise free. Each question was mapped to a variable, although the set up was able to detect the contradictory answers. We attached a few variables along with the short descriptions the full dataset and the questions may be delivered on request.

\subsection*{Examples of Some Variables:} We describe below a few representative variables used in this study. The corresponding survey questions of these variables are at: \url{https://drive.google.com/drive/   folders/1FdtwbJ2YR5lGpWssTX5HxvFvrSOSW0q-?usp=drive_link}

\begin{itemize}
    \item \texttt{Academic\_Involvment}: Number of participations in academic competitions and research work (3-year recall).
    \item \texttt{Non\_Academic\_Involvment}: Number of Participation in non-academic competitions (3-year recall)
    \item \texttt{ChatGPT\_Important\_4\_Good\_Student}: ChatGPT use willingness when competent in the subject (Highlights productivity-focused usage, even among confident students), Using ChatGPT is allowed unconditionally.
    \item \texttt{ChatGPT\_4\_Non\_Performer}: ChatGPT use for simple but uninteresting tasks or for underperforming subjects. Using ChatGPT is allowed unconditionally.
    \item \texttt{ChatGPT\_and\_useless\_subject}: ChatGPT use if  the subject is not aligned with the professional goal – using ChatGPT is allowed provided only pass marks will be given if it is correct, irrespective of answer quality.
    \item \texttt{ChatGPT\_Opposes\_Learning}: Perceived career constraint from permitted ChatGPT use.
    \item 
    \texttt{ChatGPT\_OR\_SE}: First-choice help resource when struggling (ChatGPT vs. Search).
    \item \texttt{ChatGPT\_builds\_Concept}: Sequential chatbot interactions for conceptual understanding are better and comfortable.
    \item \texttt{ChatGPT\_Better\_than\_SE}: Preference for ChatGPT over search for generating learning examples.
    \item \texttt{ChatGPT\_Explicit\_learning}: ChatGPT is helpful for active learning of something totally new (Suggests reliance on the tool for acquiring new knowledge).
    \item \texttt{ChatGPT\_Enhancement}: Need for multimodal representations in ChatGPT (e.g., graphic, video).
    \item \texttt{ChatGPT\_Assignment}: Frequency of ChatGPT use for assignments.
    \item \texttt{ChatGPT\_Used\_in\_New\_Learning}: ChatGPT usage frequency for new topics.
    \item \texttt{ChatGPT\_Helps\_New\_Learning}: Perceived helpfulness for learning new topics

    \item \texttt{ChatGPT\_Education\_Integration}: Support for ChatGPT integration in education
\end{itemize}

\subsubsection{Ethical Considerations:}

All data used in this study was anonymized and collected via voluntary online survey. No personal identifiers were included in the dataset.  Participants were informed that their responses may be used for academic research.

\section{Methodology}

This study adopts a data-driven approach to understand student behavior regarding the use of ChatGPT in academic assignments. We develop a causal model prototype using the Extreme Gradient Boosting (XGBoost) algorithm to evaluate which student characteristics and behavioral patterns best predict generative AI usage for assignment completion. The methodology encompasses data acquisition, preprocessing, exploratory analysis, model building and training, evaluation, and interpretation.

\subsubsection{Data Acquisition}  
The raw dataset was collected via a structured self-reported survey of 388 students enrolled in Russian universities, with approximately 7\% identified as international students. The survey contained 45 fields, comprising a mixture of categorical, ordinal, and numerical variables capturing demographics, behavioral preferences, learning habits, social influences, and attitudes towards ChatGPT usage. The target variable, \texttt{ChatGPT\_Assignment}, measures self-reported frequency of ChatGPT use in assignment preparation.

\subsubsection{Data Preprocessing}  
Preprocessing involved standardizing inconsistent entries (e.g., country or institution names) and removing duplicate or irrelevant columns such as timestamps and IDs, which held no predictive value and risked introducing noise. Missing values were imputed based on logical defaults or domain knowledge—for instance, categorizing unspecified African countries under an \textit{International} label.

The target variable, \texttt{ChatGPT\_Assignment}, originally recorded on a five-point Likert scale—\textit{Never}, \textit{Rarely}, \textit{Sometimes}, \textit{Frequently}, and \textit{Almost all the time}—was mapped to integers 0 through 4, preserving the ordinal nature. A novel aspect of our methodology was the formulation of two distinct classification tasks through strategic partitioning of this response variable:

\begin{itemize}
    \item \textbf{Multiclass classification}, utilizing the full five-level ordinal scale (0–4), allowing the model to predict detailed usage frequency levels.
    \item \textbf{Binary classification}, by binarizing the response variable such that values \(\geq 2\) (corresponding to \textit{Sometimes} and higher) were labeled as active ChatGPT users (class 1), \& values below 2 as rear or non-users (class 0).
\end{itemize}

This dual-task formulation enabled comparative analysis of predictive patterns under different granularities of usage behavior and constitutes a methodological novelty that offers greater insight into AI adoption behavior.

Rows with null or ambiguous target responses were removed to maintain label integrity.

Categorical features were encoded as follows:

\begin{itemize}
    \item \textbf{Ordinal variables} (e.g., \texttt{ChatGPT\_Better\_than\_SE}) were label-encoded using \texttt{LabelEncoder} from \texttt{scikit-learn}, preserving their inherent order.
    \item \textbf{Nominal variables} (e.g., \texttt{Gender}, \texttt{Faculty}) were either label-encoded or one-hot encoded, as appropriate, given tree-based models’ insensitivity to numeric ordering.
\end{itemize}

\subsubsection{Exploratory Data Analysis (EDA)}  
EDA focused on leakage detection, association strength, and predictive power of variables. Leakage detection confirmed no predictor contained information directly revealing the target (all accuracy scores < 0.5; highest: \textit{ChatGPT\_Helps\_New\_Learning} at 0.423).

Association strength, measured via Cramér’s V, identified \textit{ChatGPT\_OR\_SE} (0.501) as the top variable.  \textit{ChatGPT\_Used\_in\_New\_Learning} (0.468) and  \textit{ChatGPT\_4\_Non\_Performer} (0.481) also exhibit high scores indicating strong relevance of ChatGPT-related features. Social media and demographic variables showed weak or negligible associations.

Predictive power assessed by Information Value (IV) echoed these results: \textit{ChatGPT\_Used\_in\_New\_Learning\_Encoded} (IV = 0.385) was the strongest predictor, followed by \textit{ChatGPT\_OR\_SE\_Encoded} (0.188) and \\ \textit{ChatGPT\_Important\_4\_Good\_Student\_Encoded} (0.175). Variables with IV < 0.1 (e.g., social media usage) were excluded from modeling.

\subsubsection{Model Training}  
The dataset was partitioned into training (75\%) and test (25\%) subsets using stratified sampling to preserve class distributions, with a fixed random seed to ensure reproducibility. Stratification was critical in maintaining balanced representation across the usage classes for both classification tasks.

While scaling is generally not required for tree-based algorithms like XGBoost, all features were standardized to maintain compatibility with other pipeline components and potential baseline models (e.g., logistic regression) used for comparative evaluation.

The primary model employed was the XGBoost classifier, trained with the \texttt{binary:logistic} objective for binary classification, and the corresponding multiclass objective for the ordinal task. Hyperparameter tuning identified an optimal learning rate of 0.153, maximum tree depth of 2, 10 estimators, and subsample ratio of 0.8. The histogram-based tree construction method (\texttt{tree\_method= hist}) combined with GPU acceleration (\texttt{device= cuda}) enhanced training efficiency.

\subsubsection{Validation and Hyperparameter Optimization}  
Model optimization and validation were performed using \texttt{GridSearchCV} with 10-fold stratified cross-validation, focusing on maximizing F1-score to balance precision and recall amid class imbalance. Performance metrics demonstrated strong consistency across folds, affirming the model’s generalizability.

\subsubsection{Model Interpretation}  
To enhance transparency, intrinsic feature importance scores from XGBoost and SHAP values computed via \texttt{TreeExplainer} were used to interpret model predictions at global and instance levels. Visualization of SHAP summary and bar plots elucidated how specific features influenced the model’s outputs, enabling informed insights into student behavior factors driving generative AI adoption.

\subsubsection{Evaluation Metrics}  
Model evaluation employed standard classification metrics — accuracy, precision, recall, and F1-score — providing a comprehensive performance assessment tailored for moderate class imbalance in both binary and multiclass scenarios.

\section{Results and Discussion}

This section includes the results obtained from EDA and subsequently presents the performance of the XGBoost model in predicting student usage of ChatGPT for academic assignments and discusses the key behavioral and demographic patterns uncovered through the analysis. In this study, two predictive modeling tasks were undertaken: a \textbf{binary classification} task and a \textbf{multi-class classification} task, both implemented using the XGBoost algorithm. These models were developed to classify structured data and were evaluated through multiple interpretability metrics to ensure both performance and transparency.

\subsubsection{EDA Results:}

\begin{figure}[htbp]
    \centering
    \includegraphics[width=0.45\textwidth]{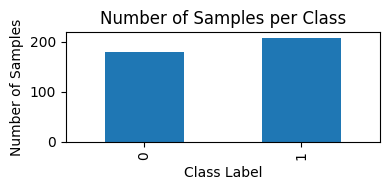}
    \includegraphics[width=0.45\textwidth]{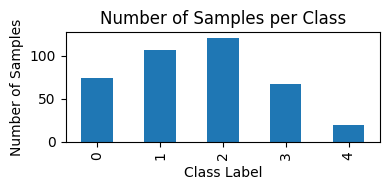}
    \caption{Left: Binary Class Distribution. Right: Multi-Class Distribution}
    \label{fig:class_count}
\end{figure}

The target variable "Chatgpt\_Assignment" has two classes for a binary classification model predicting generative AI usage in assignments. Class 0 ("Never" or "Rarely") contains 180 records, while Class 1 ("Sometimes," "Frequently," or "Almost Always") includes 208 records (fig~\ref{fig:class_count}). On the other hand, the dataset has been split into five classes according to its five unique values for multiclass classification modeling. Fig~\ref{fig:class_count} represents the distribution of data over five classes. In multiclass, the dataset is found to be imbalanced. For class 0 (Never) contain 74 instances, class 1 (Rarely) has 106, class 2 (Sometimes) 121, class 3 (Frequently) 67 and class 4 (Almost all time) 20.

\begin{table}
\centering
\caption{Comparative Model Performance Metrics}
\label{tab:comparison}
\begin{minipage}{0.48\textwidth}
\centering
\subcaption{Scores for Binary Classification}
\label{tab:metrics_a}
\begin{tabular}{|l|c|c|c|}
\hline
\textbf{Metric} & \textbf{CV} & \textbf{Training} & \textbf{Test} \\
\hline
F1 Score & 0.8015 & 0.8195 & 0.8002 \\
Accuracy & 0.8036 & 0.8230 & 0.8010 \\
Precision & 0.8082 & 0.8340 & 0.8018 \\
Recall & 0.7953 & 0.8056 & 0.7986 \\
\hline
\end{tabular}
\end{minipage}
\hfill
\begin{minipage}{0.48\textwidth}
\centering
\subcaption{Scores for Multiclass Classification}
\label{tab:metrics_b}
\begin{tabular}{|l|c|c|c|}
\hline
\textbf{Metric} & \textbf{CV} & \textbf{Training} & \textbf{Test} \\
\hline
Weighted F1 & 0.5646 & 0.6802 & 0.6450 \\
Accuracy & 0.5650 & 0.6813 & 0.6450 \\
Weighted Prec. & 0.5700 & 0.6808 & 0.6459 \\
Weighted Rec. & 0.5650 & 0.6813 & 0.6450 \\
\hline
\end{tabular}
\end{minipage}
\end{table}

\begin{table}
\centering
\caption{Binary Classification Comparison: XGBoost vs. Logistic Regression}
\label{tab:baseline_binary}
\begin{minipage}{0.48\textwidth}
\centering
\subcaption{Scores for Binary Classification}
\label{tab:metrics_binary}
\begin{tabular}{|l|c|c|}
\hline
\textbf{Metric} & \textbf{XGBoost} & \textbf{Logistic Regression} \\
\hline
Accuracy & 0.8010 & 0.7900 \\
F1 Score & 0.8002 & 0.7900 \\
Precision & 0.8018 & 0.7900 \\
Recall & 0.7986 & 0.7900 \\
\hline
\end{tabular}
\end{minipage}
\end{table}

\subsubsection{Baseline Model Comparison:}

To contextualize the performance of XGBoost classifier, we added a baseline model using Logistic Regression on the same preprocessed dataset \& train-test split. This model provides a reference to evaluate the added value of using a non-linear tree-based ensemble approach Table~\ref{tab:baseline_binary}.

The Logistic Regression model achieved an accuracy and F1-score of 79\%, slightly below the performance of XGBoost. This modest improvement suggests that while linear models capture a substantial portion of the predictive signal, the tree-based approach adds value by modeling non-linear feature interactions. Furthermore, XGBoost’s compatibility with SHAP-based interpretability provides deeper insight into behavioral patterns, justifying its use for educational analysis and policy implications.

\subsubsection{Model Performance:}

The binary classification model demonstrated robust generalization capabilities. After training and hyperparameter tuning, the model achieved good scores for several evaluation metrics (Table~\ref{tab:metrics_a}) confirming the model’s stability and minimal overfitting. The multi-class classification model, targeting five distinct classes, also demonstrated competitive performance. Following hyperparameter tuning (maximum depth = 2, learning rate = 0.153, and \texttt{reg\_lambda} = 3), the model reached an \textbf{overall accuracy of approximately 68\%}, with a \textbf{weighted F1-score of 0.68} and a \textbf{macro-averaged F1-score of 0.66}. While Classes 1, 2, and 4 showed balanced precision-recall trade-offs (F1 $>$ 0.65), Class 0 exhibited relatively lower recall ($\sim$0.54), due to imbalance data ( it provided informative stratification across the full spectrum of
user behavior) and overlapping class boundaries which introduced more complex interactions and reduced recall in certain groups (e.g., Class 0,  Table~\ref{tab:metrics_b}). Scarcity of data and presence of noise are other considerable factors in this case.

\subsubsection{SHAP Analysis \& Gain for Binary Class Classifications:} The SHAP summary plots revealed the top 15 features influencing the model's output. The most impactful variable was \texttt{ChatGPT\_Used\_in\_New\_Learning} (mean $|$ SHAP value  $| = 0.500$), indicating that students who frequently use generative AI tools for learning new topics are substantially more likely to use it for assignment preparation as well. Following this, \texttt{ChatGPT\_4\_Non\_Performer} ($0.335$), which reflects use of ChatGPT in subjects of lesser interest or performance. The study reflects strong willingness to use generative AI tools despite subject competency which allows \texttt{ChatGPT\_Important\_4\_Good\_Student} to achieve a high score ($0.207$). \texttt{ChatGPT\_Explicit\_learning}  \& \texttt{ChatGPT\_Education\_Integration} appears with the scores $0.131$ \& $0.056$ respectively.

Several demographic and behavioral variables such as \texttt{Faculty}, \texttt{Use\_X}, and \texttt{Use\_TikTok} showed lower but non-negligible importance, while features like \texttt{Sex}, \texttt{Institution}, and \texttt{Country} had near-zero SHAP values, indicating minimal influence in the predictive model.

The gain-based analysis from the XGBoost model corroborated the SHAP findings. The top three features are \texttt{ChatGPT\_Used\_in\_New\_Learning} (20.30), 
\texttt{ChatGPT\_Education\_Integration} (11.42) \& 
\texttt{ChatGPT\_4\_Non\_Performer} (10.40).

These were followed closely by \texttt{ChatGPT\_Important\_4\_Good\_Student} (8.78) and \texttt{ChatGPT\_Explicit\_learning} (8.23), confirming their consistent relevance across both interpretability techniques.

\begin{figure}[htbp]
    \centering
    \begin{minipage}{0.48\textwidth}
        \centering
        \includegraphics[width=\linewidth,height=6cm,keepaspectratio]{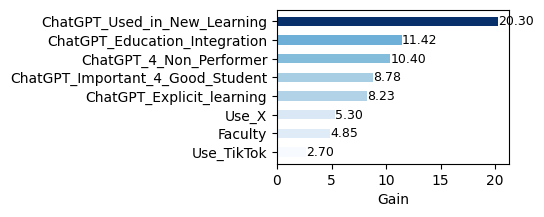}
        \caption{Feature Importance Based on XGB Gain for Binary Classification}
        \label{fig:shap_binary}
    \end{minipage}
    \hfill
    \begin{minipage}{0.48\textwidth}
        \centering
        \includegraphics[width=\linewidth,height=6cm,keepaspectratio]{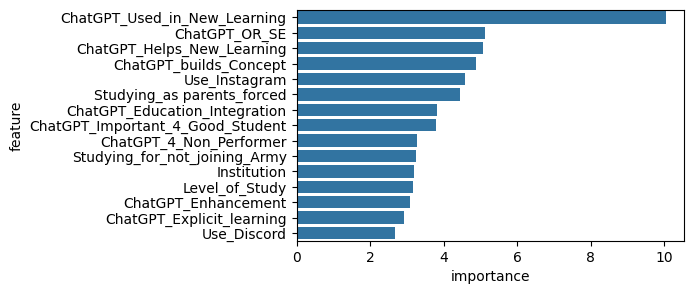}
        \caption{Feature Importance Based on XGB Gain for Multiclass Classification}
        \label{fig:shap_multi}
    \end{minipage}
\end{figure}

Our findings suggest that the propensity of students to use ChatGPT for assignment preparation is closely tied to their broader engagement with the tool as a learning facilitator with the binary classification which undergoes with validation by changing the partitions of the response variable for a multiclass study. The highest ranked variable, \texttt{ChatGPT\_Used\_in\_New\_Learning}, points to a strong behavioral alignment—students who actively explore new topics using generative AI are more likely to leverage it for task completion.

\begin{figure}[htbp]
    \centering
    \includegraphics[width=0.45\textwidth]{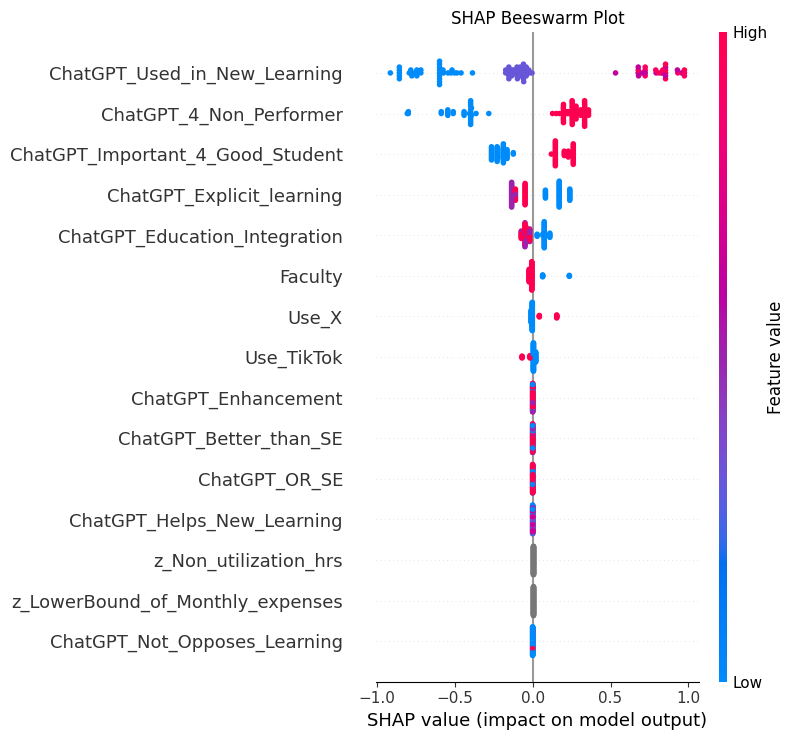}
    \includegraphics[width=0.45\textwidth]{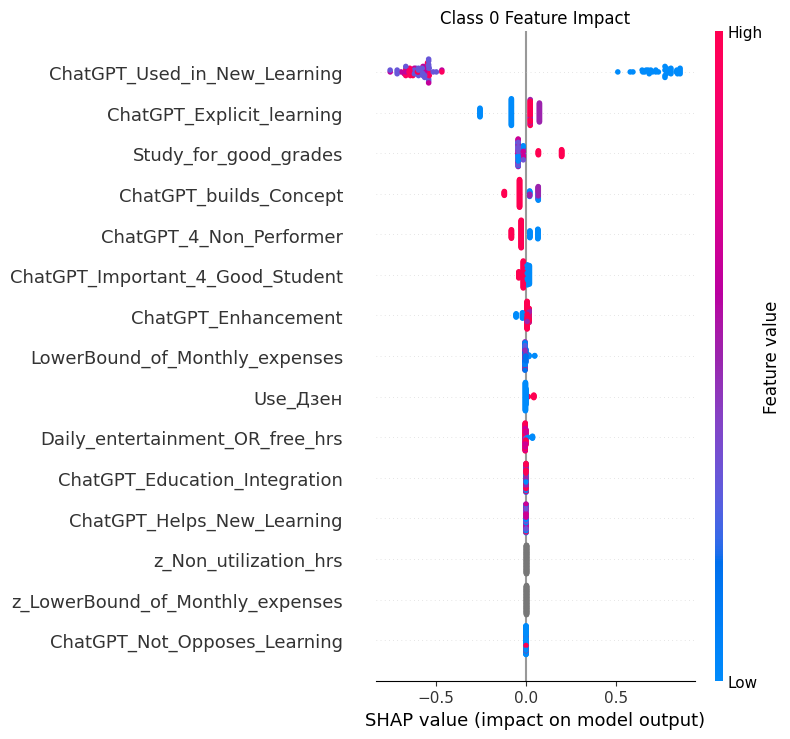}

    \caption{Beeswarm Plots: (Left) Binary Classification.  (Right) Class 0 in multiclass.}
    \label{fig:shap_plots1}
\end{figure}

The influence of \texttt{ChatGPT\_4\_Non\_Performer} underscores a pragmatic or even strategic dimension to usage: students appear more likely to rely on ChatGPT when their interest or ability in a subject is lower. This aligns with the notion of AI acting as a compensatory resource, particularly when motivation or comprehension is lacking.

Interestingly, \texttt{ChatGPT\_Important\_4\_Good\_Student} ranked third in SHAP and gain-based measures as well, suggesting that even high-performing or confident students endorse the use of ChatGPT, potentially to optimize time or enhance productivity. This challenges the traditional narrative that generative AI is primarily used to ``catch up'' and instead positions it as a legitimate academic aid.

Support for explicit learning, conceptual understanding, and educational integration also emerged as influential. The finding that support for integrating ChatGPT into education is a predictor (even if not among the top three) reflects an increasing normalization of AI tools in academic environments. The low importance of demographic variables (e.g., sex, country, level of study) further indicates that usage behavior transcends demographic boundaries, reinforcing the role of cognitive style, attitudes toward technology, and pedagogical needs as the primary drivers.

From a policy and instructional design perspective, this suggests that blanket restrictions or permissive policies may overlook the nuanced motivations behind AI tool use. For example, students may use ChatGPT not out of academic dishonesty but due to genuine needs in understanding or interest alignment. This calls for a framework that differentiates responsible AI use from misuse, perhaps by integrating AI literacy into curricula and setting clear expectations around its use.

\begin{figure}[htbp]
    \centering
    \includegraphics[width=0.45\textwidth, height=6cm]{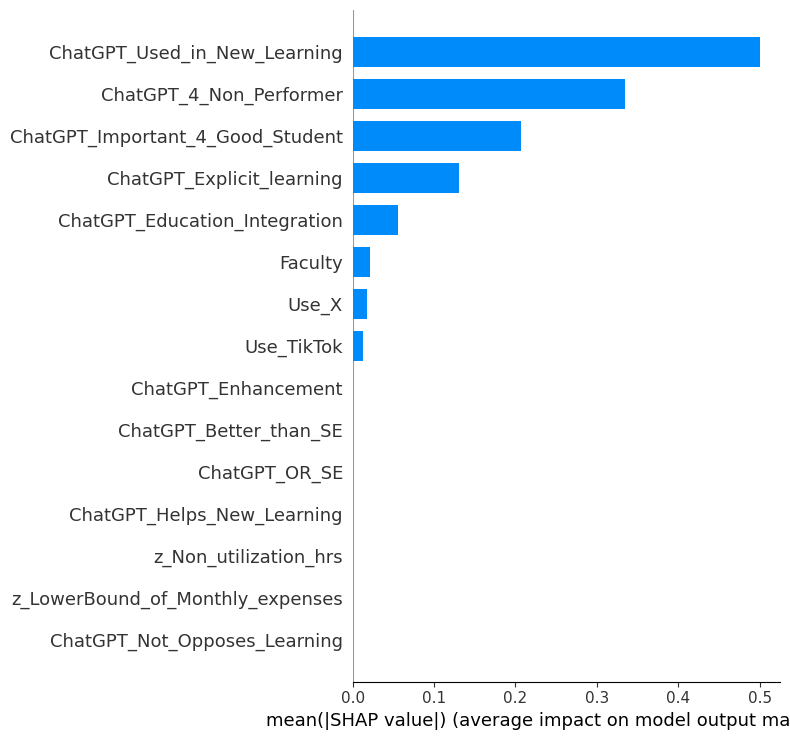}
    \includegraphics[width=0.45\textwidth, height=6cm]{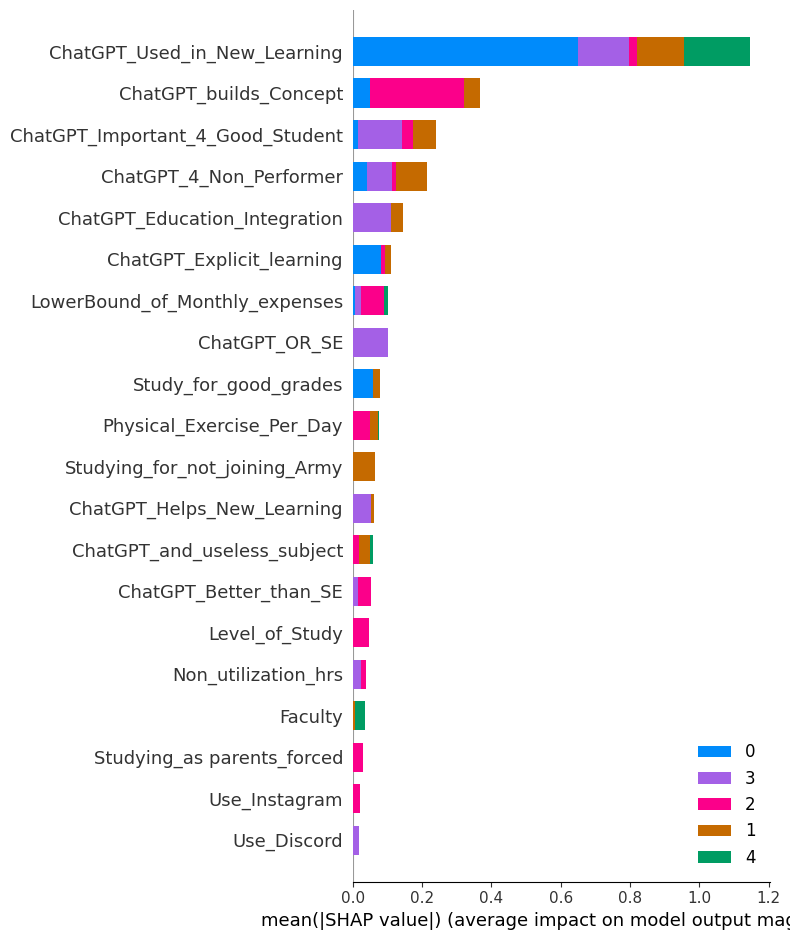}
    \caption{SHAP Feature Importance (Left) Binary Class. (Right) Multiclass}
    \label{fig:shap_plots2}
\end{figure}

\subsubsection{SHAP \& Gain Analysis for Multi Class Classifications:}The model’s gain-based feature importance and SHAP global summaries consistently highlighted \texttt{ChatGPT\_Used\_in\_New\_Learning} as the most influential predictor of assignment-related ChatGPT usage across all classes. 

This finding suggests that students who actively engage with ChatGPT for learning new topics are more inclined to utilize it for assignments, reflecting the critical role of experiential familiarity and perceived utility in shaping tool adoption.

\begin{figure}[htbp]
    \centering
    \includegraphics[width=0.45\textwidth]{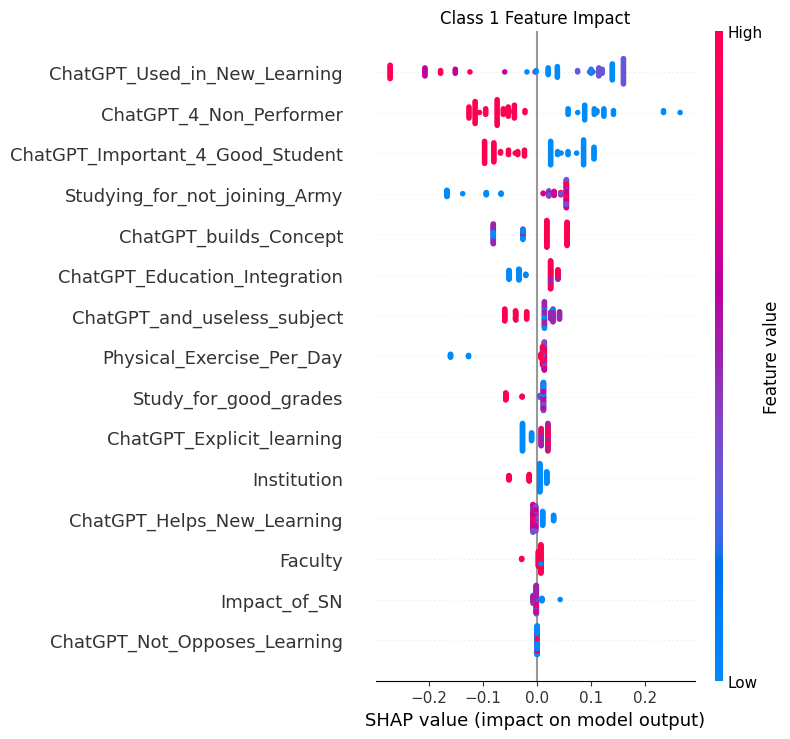}
    \includegraphics[width=0.45\textwidth]{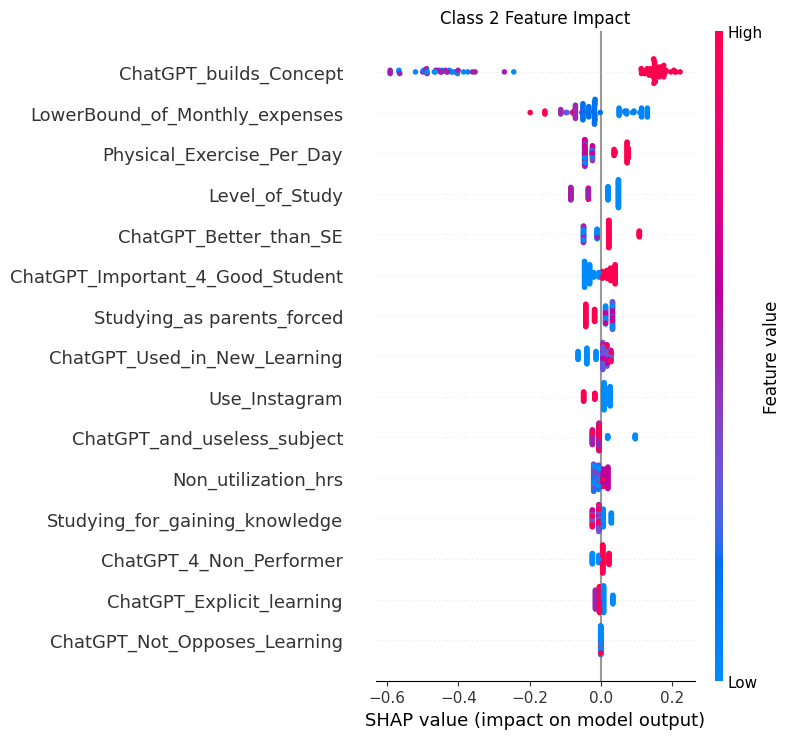}

    \caption{Beeswarm Plots: (Left) Class 1.  (Right) Class 2. Both for Multiclass.}
    \label{fig:shap_plots3}
\end{figure}

\begin{figure}[htbp]
    \centering
    \includegraphics[width=0.45\textwidth]{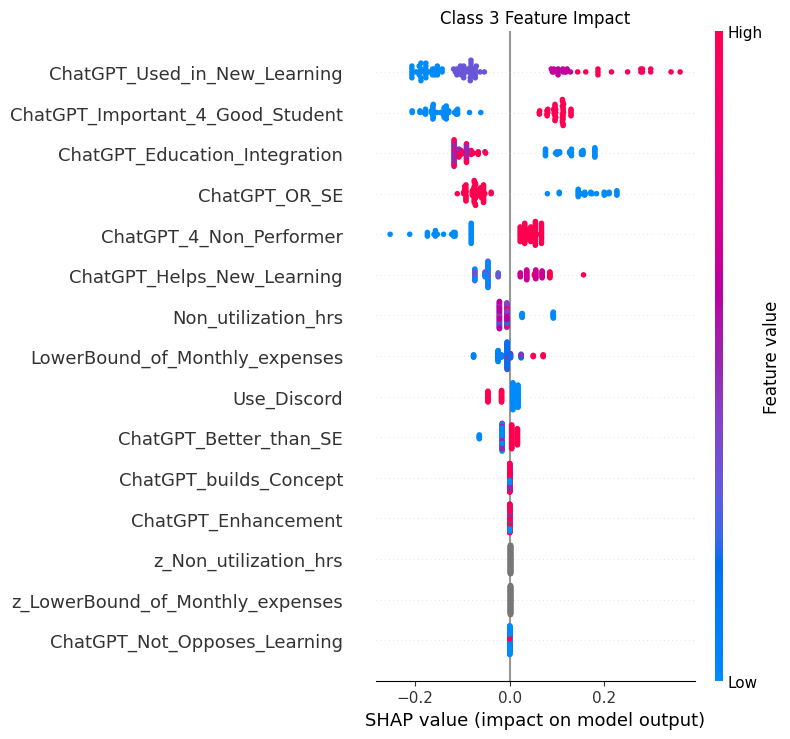}
    \includegraphics[width=0.45\textwidth]{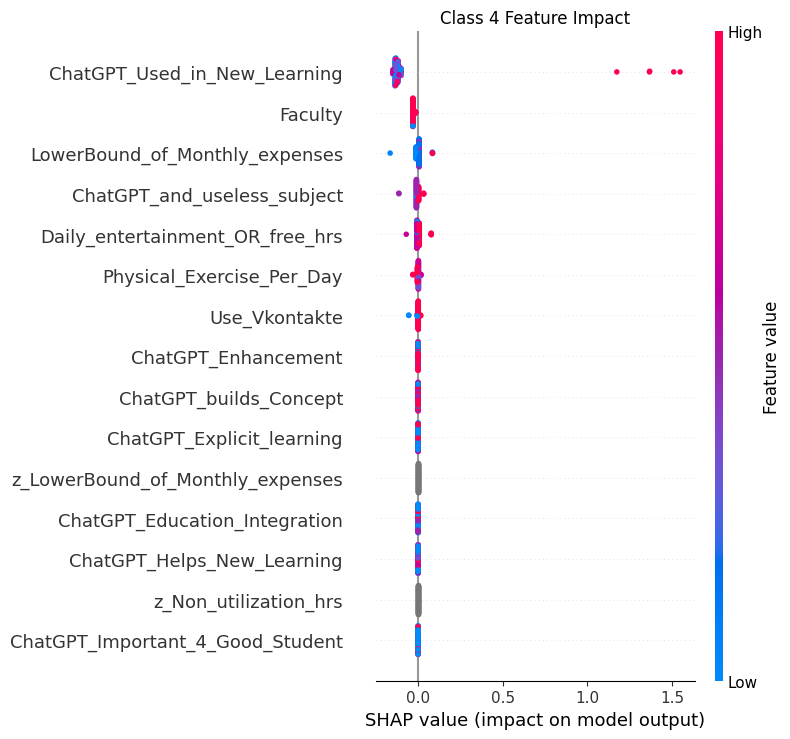}
    
    \caption{Beeswarm Plots: (Left) Class 3.  (Right) Class 4. Both for Multiclass.}
    \label{fig:shap_plots4}
\end{figure}

Other globally important features include \texttt{ChatGPT\_OR\_SE} (preference for ChatGPT over traditional search engines), \texttt{ChatGPT\_Helps\_New\_Learning}, and \texttt{ChatGPT\_builds\_Concept}, indicating that positive attitudes towards ChatGPT’s learning support capabilities strongly correlate with higher usage frequencies. These insights align with existing literature emphasizing the importance of perceived usefulness and ease of integration of digital tools in learning processes \citep{davis1989,venkatesh2012}.

\subsubsection{Class-Specific Feature Effects}

Analysis of SHAP values per usage class revealed nuanced patterns:

\begin{itemize}
    \item \textbf{Non-users (Class 0):} are primarily characterized by low experiential engagement with ChatGPT in learning new topics, as reflected by the dominant influence of \texttt{ChatGPT\_Used\_in\_New\_Learning} with minimal contributions from other variables (fig~\ref{fig:shap_plots1}). This suggests a lack of familiarity or trust in the tool as a learning aid, consistent with earlier findings that unfamiliarity impedes technology adoption \citep{rogers2003}.
    
    \item \textbf{Rare users (Class 1):} exhibit moderate reliance on ChatGPT, influenced by factors such as perceived utility in non-performing or less interesting subjects (fig~\ref{fig:shap_plots3})(\texttt{ChatGPT\_4\_Non\_Performer}) and conditional acceptance of ChatGPT use (\texttt{ChatGPT\_Important\_4\_Good\_Student}). This group may use ChatGPT reactively rather than proactively.
    
    \item \textbf{Moderate users (Class 2):} display higher importance of conceptual understanding facilitated by ChatGPT (fig~\ref{fig:shap_plots3})(\texttt{ChatGPT\_builds\_Concept}) and show sensitivity to socio-economic factors like monthly expenses, possibly reflecting access or resource considerations. Their moderate usage implies selective integration of ChatGPT into their study practices.
    
    \item \textbf{Frequent users (Class 3)} and \textbf{almost always users (Class 4):} demonstrate strong positive associations with both experiential use in learning new material and attitudinal variables supporting ChatGPT’s educational integration. Notably, the almost always users’ reliance on ChatGPT appears habitual and less influenced by external factors, indicating potential dependency or ingrained usage patterns (fig~\ref{fig:shap_plots4}).
\end{itemize}

\subsubsection{Interpretation and Educational Implications:}

The analysis confirms a behavioral continuum in ChatGPT adoption, shaped more by cognitive orientation and learning strategies than by demographics. The convergence of SHAP and gain importance highlights \texttt{ChatGPT\_Used\_in\_New\_Learning} as the strongest predictor, emphasizing the role of experiential familiarity in assignment-level AI use. Variables such as \texttt{ChatGPT\_4\_Non\_Performer} and \\ \texttt{ChatGPT\_Important\_4\_Good\_Student} suggest compensatory and enhancement-driven motivations. These findings imply that students engage with ChatGPT both as a conceptual aid in difficult subjects and as a productivity tool in familiar ones. Educational practices should promote responsible, structured use—raising awareness among non-users and guiding frequent users toward ethical engagement. Additionally, socio-economic factors like parental pressure and study expenses point to broader contextual influences, underscoring the need for holistic, differentiated strategies in AI-integrated learning environments.

\section{Conclusion}

This study applied interpretable machine learning to examine behavioral and attitudinal predictors of ChatGPT use in academic assignments among university students. Analysis of survey responses from 388 students using XGBoost and SHAP revealed that experiential familiarity with ChatGPT—especially for learning new topics—was the strongest predictor of assignment-related usage. In contrast, demographic factors such as gender, institution, or country showed minimal influence.

Findings indicate a behavioral continuum of adoption, shaped by motivations ranging from conceptual support and efficiency to strategic use in weaker subjects. This challenges the view of AI use as purely opportunistic, highlighting the need for nuanced, AI-aware assessment policies.

Despite strong predictive performance, the study has limitations. It relied on cross-sectional, self-reported data, which may carry social desirability or recall bias. The sample was predominantly Russian, limiting broader generalizability. In the multiclass classification task, no SMOTE or synthetic resampling was applied—though stratified sampling was used to reduce class imbalance, recall for minority classes may still be suboptimal.

Future work should incorporate longitudinal designs and qualitative methods to uncover causal and contextual factors behind AI use. Applying augmented or balanced datasets, and exploring alternative models (e.g., deep or ensemble learning), could enhance performance, particularly in imbalanced settings. Interactive SHAP-based dashboards may support educators in tracking AI engagement, while controlled interventions or A/B testing could guide policy on AI-integrated assessment.

Ultimately, this work offers actionable, data-driven insights to support ethical and effective integration of generative AI into education.

\begin{credits}
\subsubsection{\ackname}This research was carried out as part of an internal initiative at Educational and Scientific Center
for Humanities and Social
Sciences of MIPT (Moscow Institute of Physics and Technology).

\subsubsection{\discintname}
The authors declare that they have no competing interests relevant to the content of this article.acknowledgments\footnote{This publication is a preprint and has not been peer-reviewed. It is made available under the selected license. Any unauthorized use or violation of the license terms is strictly prohibited and will be the responsibility of the individual or entity engaging in such use.},
\end{credits}
%
%
%
%


\bibliographystyle{plain}

\end{document}